\documentclass[9pt,twocolumn,twoside]{opticajnl} 
\journal{opticajournal} 

\setboolean{shortarticle}{false}

\usepackage{lineno}

\usepackage{gensymb}
\usepackage{geometry, siunitx, enumitem,  comment, wasysym} 
\usepackage{tocloft} 
\usepackage[version=4]{mhchem}
\usepackage{xurl}

\title{Laser, Vacuum, and Gas Reaction Chamber for \emph{Operando} Measurements at NSLS-II's 28-ID-2}
\author[a,b]{Lauren Y. Moghimi}
\author[a,b,+]{Patrik Johansson}
\author[b,c,+]{Subhechchha Paul}
\author[a,b,c,+]{Yifan Wang}
\author[d,e]{Sara Irvine}
\author[a,b]{Remington Graham}
\author[a,b]{Zane Taylor}
\author[f]{Angel A. Martinez}
\author[f]{John T. Markert}
\author[g]{John Trunk}
\author[g]{Hui Zhong}
\author[g]{Jianming Bai}
\author[g]{Sanjit Ghose}
\author[a,b,c,*]{Leora Dresselhaus-Marais}

\affil[a]{Department of Materials Science and Engineering, Stanford University, Stanford, CA 94305, USA}
\affil[b]{PULSE Institute, SLAC National Accelerator Laboratory, Menlo Park, CA 95024, USA}
\affil[c]{Department of Mechanical Engineering, Stanford University, Stanford, CA 94305, USA}
\affil[d]{SIMES, SLAC National Accelerator Laboratory, Menlo Park, CA 95024, USA}
\affil[e]{Department of Applied Physics, Stanford University, Stanford, CA 94305, USA}
\affil[f]{Department of Physics, The University of Texas at Austin, Austin, Texas 78712, USA}
\affil[g]{National Synchrotron Light Source II, Brookhaven National Laboratory, Upton, NY 11973, USA}
\affil[+]{Authors contributed equally}
\affil[*]{Correspondence email: leoradm@stanford.edu}

\begin{abstract}
We present a laser reaction chamber that we developed for \emph{in-situ/operando} X-ray diffraction measurements at the NSLS-II 28-ID-2 XPD (X-Ray Powder Diffraction) beamline. This chamber allows for rapid and dynamic sample heating under specialized gas environments, spanning ambient conditions down to vacuum pressures. We demonstrate the capabilities of this setup through two applications: laser-driven heating in polycrystalline iron oxide and in single crystal \ce{WTe_2}. Our measurements reveal the ability to resolve chemical reaction kinetics over minutes with 1-s time resolution. This setup advances opportunities for \emph{in-situ/operando} XRD studies in both bulk and single crystal materials.
\end{abstract}

\setboolean{displaycopyright}{false} 

\renewcommand{\thesection}{\arabic{section}} 
\renewcommand{\thesubsection}{\thesection\alph{subsection}}

\begin{document}

\maketitle
\tableofcontents
\section{Introduction}
Time-resolved measurements give insight into the mechanisms underlying material synthesis and processing. Chemical kinetic parameters like reaction rate constants ($k$) are used to obtain activation energies ($E_a$) in Arrhenius-type datasets, which provide insights into the reduction mechanisms and assist with process optimization. Reduction of solid-state iron oxides has been studied with \emph{in-situ} techniques including thermogravimetric analysis (TGA) and temperature-programmed X-ray diffraction (XRD) to measure phase transition temperatures \cite{bulavchenko2019influence,qu2014thermal,luu2025towards}. These works, and similar studies, have been helpful to the \ce{H_2}-based reduction field by experimentally validating transition temperatures and rate-limiting steps in varied sample systems. However, such techniques cannot capture phase distributions or reaction dynamics for changes on the order of minutes, or with non-equilibrium or \emph{operando} conditions. 

\begin{figure*}[ht]
\centering
\mbox{\includegraphics[width=0.8\linewidth]{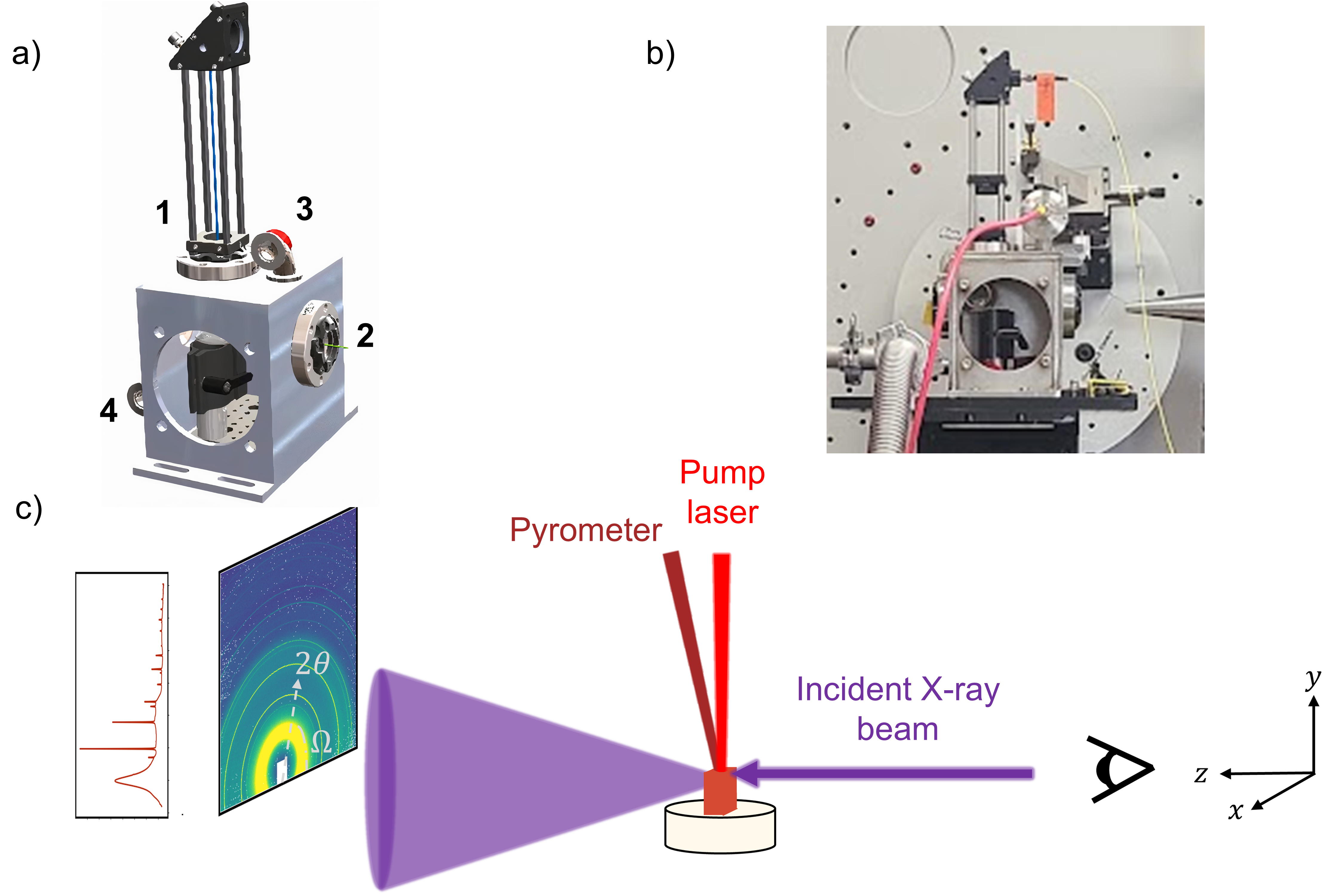}}
\caption{A (a) model and (b) picture of the reaction chamber for experiments at 28-ID-2. 1-4 label the laser viewport, X-ray entry viewport, gas inlet, and gas outlet or vacuum port, respectively. (c) Schematic of the measurement configuration for an iron oxide pellet.} 
\label{fig:setup}
\end{figure*}

\emph{In-situ} reflection-mode XRD is a widely accessible technique to measure structural changes under sample heating, but its capability is often limited to sample surfaces due to small X-ray penetration depths relative to the sample width. In comparison to reflection-mode XRD, transmission-mode XRD has the benefit of enabling either bulk or surface analysis, depending on the sample geometry and scan configuration. Inherently, the interaction volume is larger in transmission-mode XRD, allowing users to capture microstructural results that are representative of a full sample rather than just the sample surface.

The high energy X-ray (\textasciitilde70 keV) provided by the 28-ID-2 beamline at the National Synchrotron Light Source II (NSLS-II) enables single-shot transmission-mode XRD measurements with a 1-s time interval. This includes \emph{in-situ} powder XRD measurements that can be used to study phase transitions on otherwise X-ray-opaque samples \cite{shi2013performance}. The adjustable beam size can be expanded to match the scale of heterogeneity in 1 mm-sized samples, or reduced to $\SI{100}{\micro m}$ to capture the local structural evolution on a slice of the sample. To date, the beamline has supported a variety of \emph{in-situ} and \emph{operando} XRD experiments that benefit from instrumentation such as heating sources (lamp furnace, heating stage, coil heater, hot air blower) and specialized sample environments \cite{elbakhshwan2016sample,haas2021compactfurnace,avila2021phase,plonka2019effect}. The wide variety of instruments available at this beamline have impacted fields spanning battery materials, flash sintering, and catalysis. For some applications in catalysis and chemistry, however, an additional reaction chamber capable of wavelength-tunable photochemistry with variable gas environments is required. Here we present an adaptable laser reaction chamber for measurements at the 28-ID-2 beamline, under rapid/dynamic sample heating and specialized gas environments, from ambient down to vacuum pressures. To demonstrate this setup, we present its use in two areas: 1) chemical reduction in polycrystalline iron oxides and 2) phase transitions in single crystal \ce{WTe_2}. 

We show how this setup can resolve minutes-long chemical reactions, measured with \textasciitilde1-s time resolution. This setup can be useful for various \emph{in-situ / operando} applications investigating photo-induced transitions and laser-driven heating, such as photochemical catalysis, laser spot welding, and pyrometallurgical minerals processing.

\section{Design and implementation}
We designed our reaction chamber to interface with beamline 28-ID-2, enabling XRD measurements up to $2\theta=15\degree$ at 68 keV in transmission mode. Fig. \ref{fig:setup}a Fig. \ref{fig:setup}b show the model and picture of the reaction chamber.

\subsection{Beamline specifications}
The $2\theta$ detection range and $q$ resolution are two parameters that should be considered when deciding whether near-field or far-field diffraction measurements are more appropriate for experiments. We utilize the beamline's far-field detector---Perkin-Elmer 1621---to achieve high $q$ resolution. As a result, the size and position of the chamber's X-ray exit port along the sample-to-detector distance limit the $2\theta$ detection range. In experiments with the chamber at 68 keV, all phases of interest were resolvable in our mixed-phase samples and we accessed $2\theta_{max}=15\degree$, which captured up to $q_{max}=10.5$ \AA{}$^{-1}$. In terms of material parameters, down to 0.095 \AA{} that satisfy the diffraction condition can be reached by this $q$ range. As this sufficiently captures the full range of $d$-spacings for phase identification in most materials, Rietveld refinement may be used for quantitative phase analysis. This work uses the current beamline detectors that have the ability to take data at 1 shot/s, though planned detector upgrades (high-speed pixel counting) will soon support millisecond detection.

\subsection{Chamber geometry and environment}
To ensure that our chamber can operate under a variety of pressures, gas environments, and high temperatures, we place all motors, sensors, and optics outside of the chamber. The sample position relative to the X-ray beam is controlled by two mechanisms: (1) a motorized $xyz$ stage outside the chamber and (2) a manual sample mount inside the chamber. Our selected design eliminates the need for vacuum-compatible motors. 

Chamber attachments are placed at ports facing the beamline-accessible side of the chamber stage (i.e. in +$x$), in case any modifications are needed while the chamber is mounted. KF16 flange elbows are welded onto the chamber for the gas inlet and outlet---enabling an adaptable setup---such as switching between gas and vacuum inputs, or routing exhaust from the chamber. An optional digital vacuum gauge can measure the pressure within the chamber, and may be placed or removed at the exit elbow. 

Our chamber reached vacuum pressures as low as 0.04 Torr. To appropriately isolate the effect of specific gas atmospheres on the iron oxide reduction behavior, we evacuated the chamber near this minimum pressure before introducing either 3\% \ce{H_2} in Ar or 100\% Ar at a flow rate of 5 standard liters per minute (slm).

\subsection{Sample mounting}
Sample size is selected based on the X-ray attenuation vs. integration volume of interest. For example, to study collective phenomena in a bulk sample, a large sample and wide beam size may be preferable to uniformly integrate heterogeneous processes. We opted for $< \SI{2}{mm}$ width powder samples for sufficient transmission of all phases that emerged during reduction from \ce{Fe_2O_3} (hematite). Our calculations showed that full-density pure Fe pellets at this width would have adequate transmissivity (23\%) at 68 keV \cite{NISTattxray}. To capture bulk reaction dynamics, we produced smaller $\SI{1}{mm}$ cubic samples to fit entirely within an expanded beam of $\SI{1.2}{mm} \times \SI{1.5}{mm}$. Surfaces could also be studied using the same experimental configuration by restricting the overlap between the sample and X-ray to the surface.

We tested three forms of solid samples with this setup:  a loose powder, pelletized powder, and sheet of single crystal. Loose powder samples were mounted in rectangular cuvettes to contain the samples in rigid form. Fused quartz cuvettes with internal width of 2 mm and wall thickness of 0.75 mm were chosen to minimize parasitic contributions to the XRD signal. The cuvettes were positioned upright, such that the sample was directly exposed to the laser beam from above and that gas could pass over the sample freely (Fig. \ref{fig:setup}c). 

We also prepared pellets of the hematite powder to reduce the risk of sample displacement away from the laser-illuminated spot and X-ray collection area, which would likely occur for powders due to violent \ce{O_2} off-gassing. We pre-mounted samples to dowel pins and fixed the position of the dowel holder to maintain consistent sample alignment across experiments. The dowel holder was a translating optical post assembly, which accommodated 0.635 mm of height adjustment per revolution. 

Single crystals of \ce{WTe_2} were oriented with and without a tilt to observe microstructural transitions for diffraction planes of interest (i.e. along $\Vec{c}$, Fig. \ref{fig:combined-wte2}a). Because of the high X-ray energy not previously seen in other studies, transmission through orientations besides the sample's stacking axis was possible. We adapted a 3D-printed tilt mount to a vertically-adjustable platform inside the reaction chamber and utilized two mounting geometries, where $\Vec{b}$ was either $0\degree$ or $75\degree$ with respect to the X-ray beam axis ($z$).

For all samples, both the dowel holder and platform were manually adjusted in the vertical direction ($y$) and positioned in the $xz$-plane. We fixed the mount position once we determined the desired locations during alignment. Fine control of the sample relative to the X-ray beam was achieved using a motorized $xyz$ stage outside the chamber.

\begin{figure}[ht]
\centering
\mbox{\includegraphics[width=\linewidth]{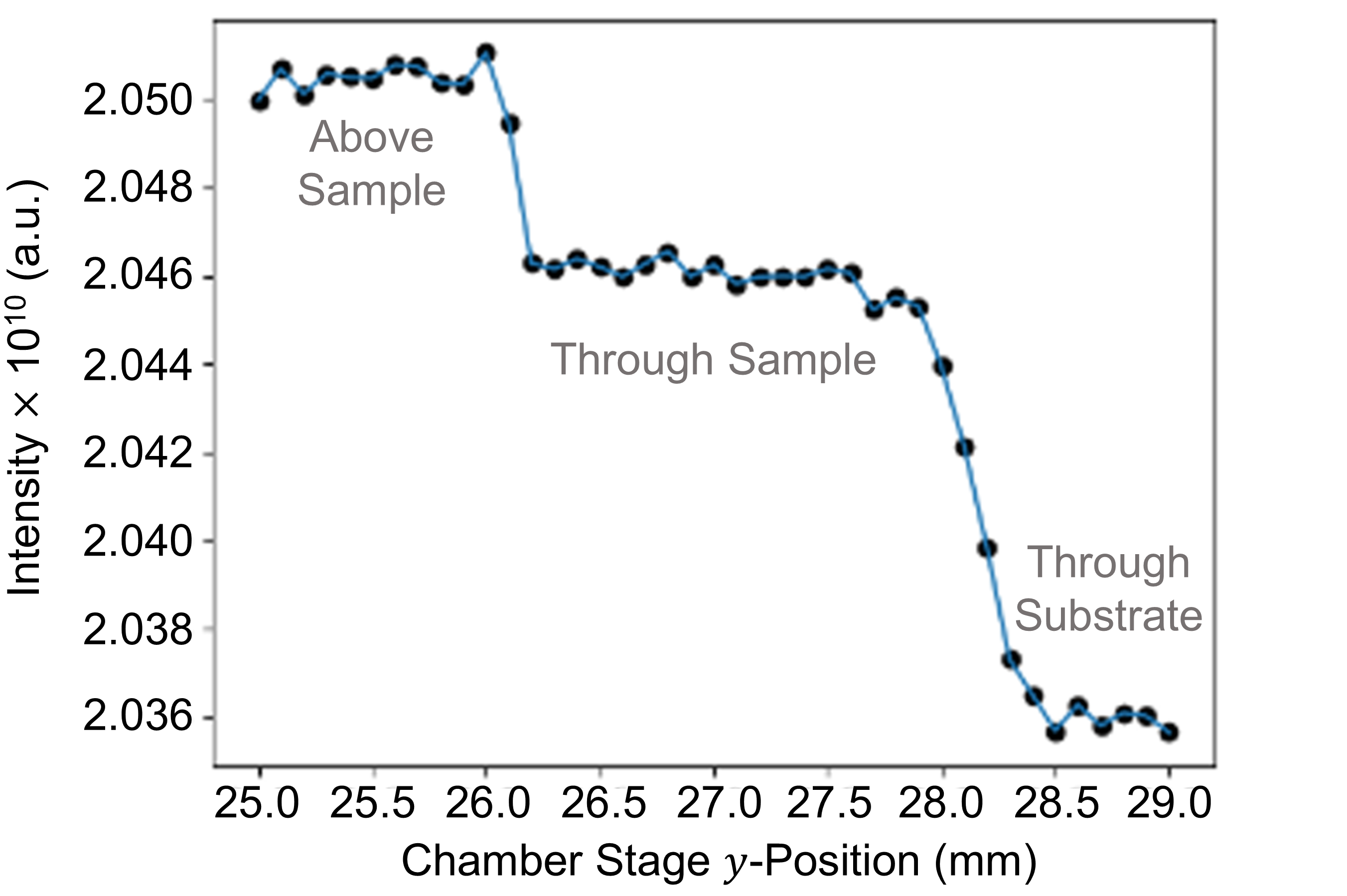}}
\caption{Transmitted X-ray intensity as a function of stage $y$-position (mm) for an iron oxide pellet. From left to right, the plateaus correspond to the X-ray above the sample, through the sample, and through the substrate.}
\label{fig:height_alignment}
\end{figure}

\subsection{Alignment}
\label{subsection:alignment}
After alignment of the upstream X-ray optics preceding the chamber, three steps were done before the setup was ready for well-calibrated measurements:
\begin{enumerate}
    \item position the chamber to the X-ray beam,
    \item align the laser beam to the X-ray beam, and
    \item focus the pyrometer to the laser and X-ray beams.
\end{enumerate}
Engineering controls and fixed sample mounting positions allowed us to maintain alignment between the 3 beams across experiments. We mounted the pyrometer and laser input to the $xyz$ stage or directly to the chamber so that the pyrometer always measured the illumination spot on the sample surface. 

In general, the chamber should be aligned with the X-ray beam in the $xy$-plane so that the measurable $2\theta$ range is maximized. The azimuthal angle along the diffraction rings captures structural mosaicity in the samples, which can be especially important for non-polycrystalline samples. As seen in Fig. \ref{fig:setup}c the 2D detector was offset above the X-ray axis so that the captured azimuthal range was just over $\Delta\Omega=180\degree$. To achieve up to $2\theta_{max}=15\degree$ with the azimuthal range as shown, we positioned the chamber by centering the X-ray viewports to the X-ray guide beam. We then adjusted the sample mount and the sample to the X-ray guide beam in $y$. We refined the chamber's alignment to the X-ray by measuring the transmitted X-ray intensity while scanning the vertical motor position, as seen in Fig. \ref{fig:height_alignment}. Before setting the chosen height for the $xyz$ stage, we checked that we could measure sufficient sample diffraction at the selected position.

To measure XRD of the sample region under laser illumination, we next aligned the laser beam in the $xz$-plane to the X-ray guide beam. Ideal alignment of the laser beam has the laser at an angle of $90\degree$ with respect to the X-ray beam, illuminating the sample on the top. For safety, we used the minimum power during laser alignment and higher power only via remote control after the hutch was unoccupied by personnel. We used a near infrared (NIR) viewing card mounted to a viewport cover to help align the NIR laser beam. We aligned and fixed the $xz$-position of the sample mount to the intersection of the laser and X-ray beams.

Finally, we aligned the pyrometer to the sample surface to measure temperature at the laser-illuminated spot. The pyrometer has two $\SI{635}{nm}$ guide beams to assist with alignment; the beams cross at the device's focal position. The collection area is given by the spot size, which is nominally $\SI{0.3}{mm}$ at the focal length ($\SI{150}{mm}$). The pyrometer was positioned at a grazing angle, which increased the collection area on the sample surface. Beyond alignment, we discuss the pyrometer specifically in Section \ref{subsection:T_measurements}.

\subsection{Laser integration}
\begin{figure}[ht]
\centering
\mbox{\includegraphics[width=0.8\linewidth]{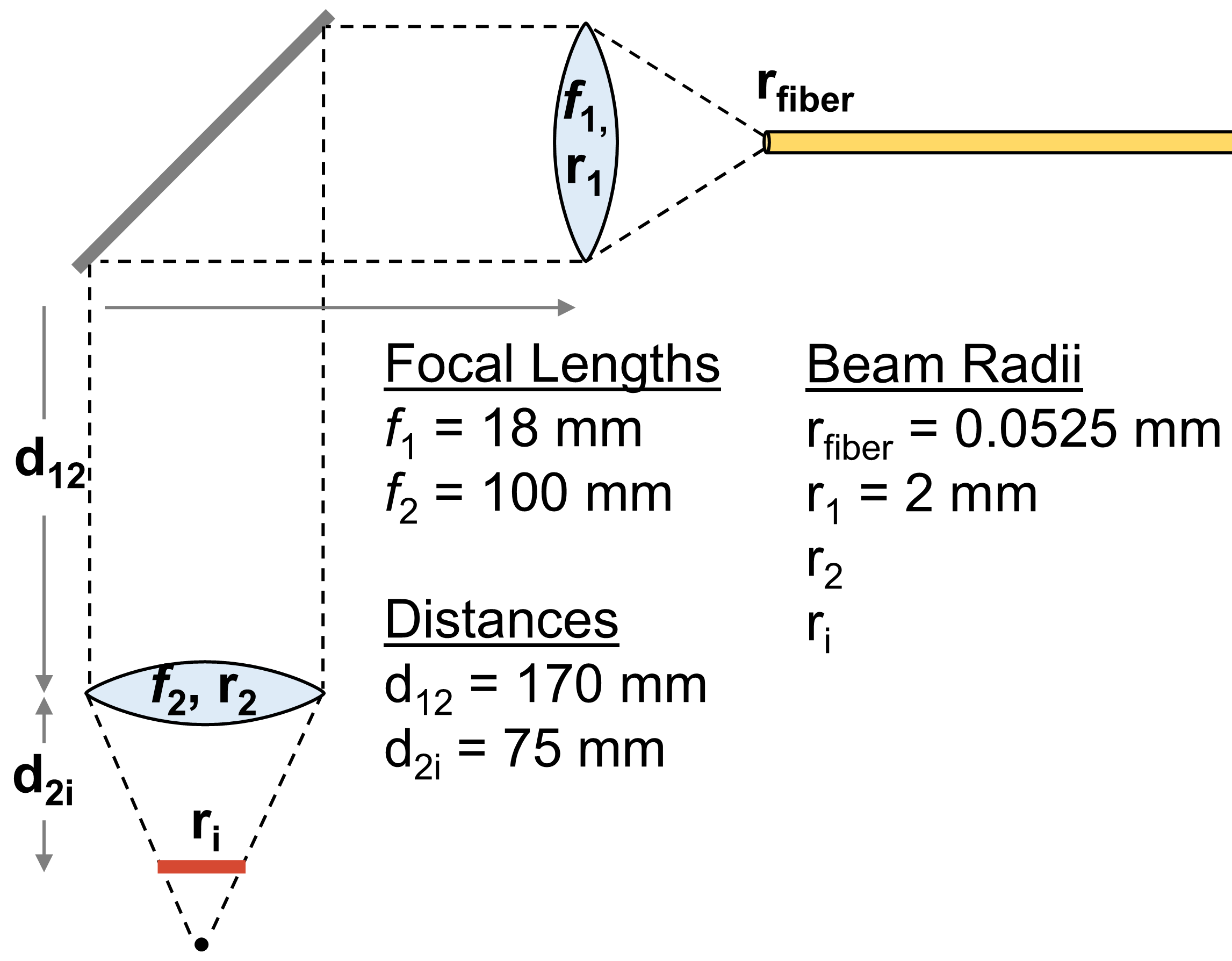}}
\caption{Schematic of the optics along the laser beam path, used to calculate the incident spot size. $r_1$ is the beam radius just after lens 1, while $r_2$ is the beam radius just before lens 2.}
\label{fig:spot_size}
\end{figure}

The laser in this setup can be used to drive photochemical or thermal phase transitions, though the samples studied in this work were limited to temperature-driven transitions. Samples were heated by exposure to a continuous-wave 10 W fiber-coupled laser. The laser beam was guided along a cage system that mounted to the chamber so that alignment was independent of chamber movement. We tuned the incident spot size using a fiber collimator and focusing lens that were compatible with the laser wavelength and using the setup shown in Fig. \ref{fig:spot_size}. The mirror was at a $45\degree$ tilt from the optical axis and had high reflectivity (\>97\%) at the laser wavelength. To maximize the laser power transmitted through the laser viewport, we chose an antireflective window that spanned the 976 nm (NIR) and 445 nm (blue) wavelengths used thus far.

The spot size at the sample surface was determined by the following equations, where $\phi/2=0.018\degree$ is the divergence half-angle from $f_1$. The beam diameter at $f_2$ is given by 
\begin{align*}
    2\cdot r_{2}&=2\cdot( r_{1}+d_{12}\cdot\tan(\phi/2)) \\
    &= \SI{4.1}{mm}.
\end{align*}
Similarly, the incident beam diameter is given by
\begin{align*}
    2\cdot r_i&=\frac{2\cdot r_{2}}{f_2}\cdot|d_{2i}-f_2| \\
    &= \SI{1.0}{mm}.
\end{align*}

The electrical input for the laser was remotely controlled and digitally logged using the desktop control system provided by the power supply manufacturer (BK Precision 9103). Thus, our experiments were able to track incident optical power as a function of time. We calibrated the electrical power to the optical power illuminating the sample using a powermeter. 

\subsection{Measurements of laser heating}
\label{subsection:T_measurements}
We tracked the sample temperature remotely and \emph{in situ} using a pyrometer (Micro-Epsilon CTLM-3H3CF2-C3), which offered several advantages over thermocouple feedthrough measurements. Optical pyrometer measurements were preferred in this work so that physical contact with the sample was not necessary. Non-invasive temperature measurements offer opportunities to study processing environments that may degrade thermocouples, such as chemically-reactive, high-temperature, or low-pressure conditions. We focused the pyrometer to the sample surface following the alignment scheme discussed in Section \ref{subsection:alignment}. We mounted a 1-$\SI{ }{\micro m}$ longpass filter to prevent interference from our NIR laser with temperature measurements. We used a \ce{CaF_2} window for the pyrometer viewport to ensure high IR transmittance ($>90\%$) up to a wavelength of $\SI{7}{\um}$.

\begin{figure*}[ht]
\centering
\mbox{\includegraphics[width=\linewidth]
{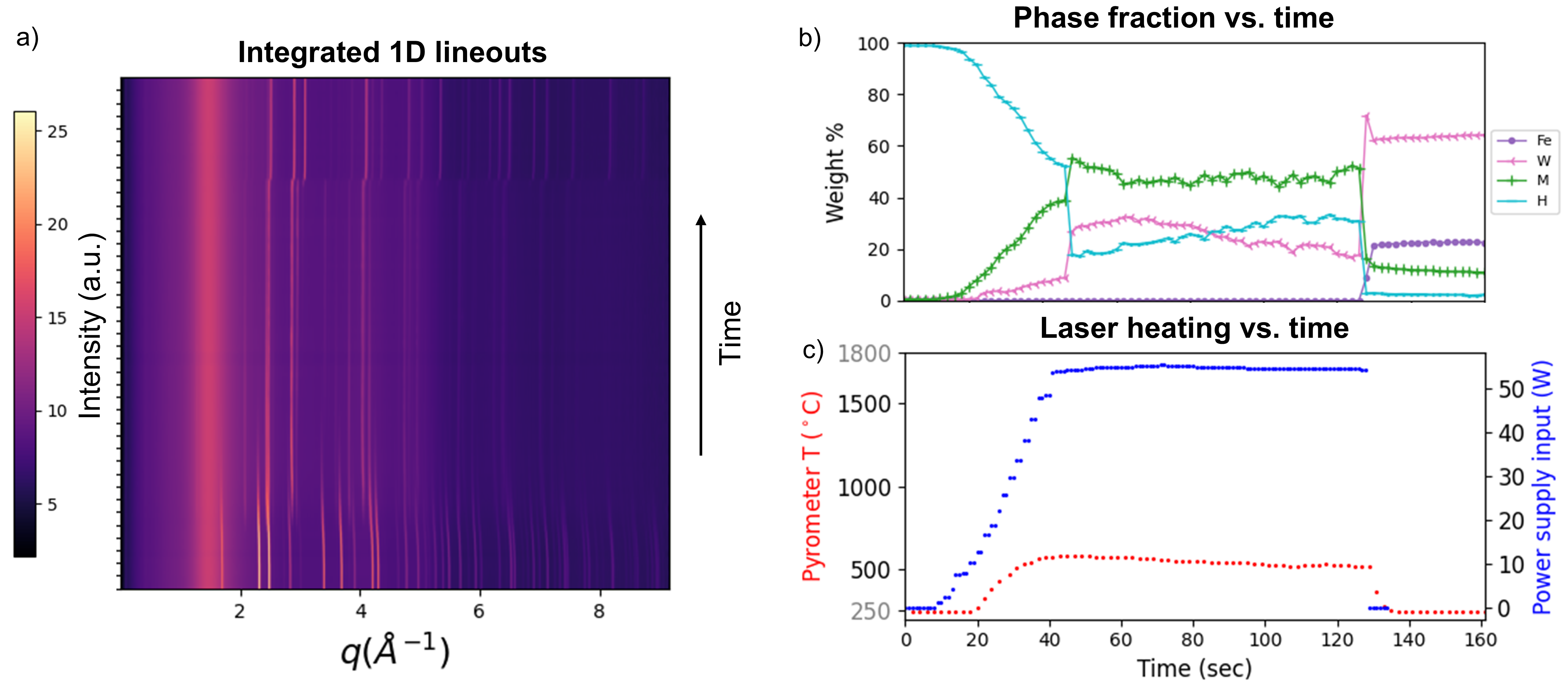}}
\caption{Data from \emph{operando} hematite experiments under 3\% \ce{H_2} and laser heating with $\lambda_{laser}=\SI{976}{nm}$ reaching up to the maximum power $P_{max}=\SI{10}{W}$ at the sample surface. (a) Waterfall plot of the integrated 1D X-ray powder diffraction vs. time. (b) Corresponding phase fraction vs. time and (c) plot of the laser input and raw pyrometer measurements. Phase distributions were analyzed with Rietveld refinement.}
\label{fig:ex_FeOx_results}
\end{figure*}

The device's temperature readouts are between 250-1800$\degree$C. To calibrate the pyrometer after measurements, we calculated the temperature of heated pellets using the shift in diffraction peak position, as quantified by 
\begin{align*}
T = T_{0}+\frac{d-d_{0}}{\alpha_L \cdot d_{0}},
\end{align*}
where $\alpha_L$ is the linear expansion coefficient, $T$ is the sample temperature, $d$ is the interplanar spacing at this temperature, and $T_0$, $d_0$ are at room temperature. We additionally prepared pellets of 33 wt\% \ce{CeO_2} powder in hematite as a calibration standard for temperature, as \ce{CeO_2} (ceria) is a NIST standard for quantitative XRD analysis. We chose this ceria-to-hematite ratio to preserve the pellet's optical and thermal properties, while ensuring sufficiently high intensity of ceria's main XRD peak within the hematite matrix.

\section{Example applications}
In this section, we discuss experiments on pellet and single crystal samples that we studied using our laser reaction chamber. We plot everything in $q$ such that our peaks are well-resolved and photon energy-invariant, for comparison to other studies and materials.

\subsection{Gas-solid reaction of pelletized hematite}
We begin by presenting a study of hematite pellet reduction in various gas environments during laser irradiation. Hematite is among the iron oxides that are established chemical- and photo-catalysts \cite{leland1987photochemistry, khedr2009synthesis, pang2016research, parkinson2016iron, shaikhutdinov1999structure}, with applications spanning water splitting \cite{badia2013water}, water oxidation \cite{kanazawa2017chromium}, and nitrate splitting \cite{jung2014development}. We used our setup to identify hematite's reduction pathway under \ce{H_2}, \ce{Ar}, and vacuum, though we focus our demonstration to

we present only \ce{H_2} as a demonstration. 

The samples reached high temperatures up to \ce{\degree C} during heating with the maximum laser power. The heating laser also enabled fast heating rates that could bring samples to the steady-state temperature within 2 seconds at constant laser power. Fig. \ref{fig:ex_FeOx_results}a and Fig. \ref{fig:ex_FeOx_results}b show the raw and processed data, respectively, for hematite pellets heated in the 3\% \ce{H_2} environment. As shown by the laser power vs. temperature plots in Fig. \ref{fig:ex_FeOx_results}c, the reactions in hematite sample exhibited three dominant stages at the times when the laser ramped, held at maximum power, and shut off (Fig. \ref{fig:ex_FeOx_results}b). 

This experiment has allowed us to observe the dynamics of iron oxide reduction in samples heated to high temperatures. TGA and temperature-programmed XRD studies have shown clear step-wise reduction following $\ce{Fe_2O_3}\rightarrow\ce{Fe_3O_4}\rightarrow\ce{FeO}\rightarrow \ce{Fe}$ \cite{jozwiak2007reduction}. These studies hence produce reaction rates for each step-wise transition, or produce the conversion degree for the overall progression towards iron \cite{spreitzer2019reduction, xing2020kinetic}. Our results show an even more complicated picture for reduction in non-equilibrium conditions (faster heating rates), wherein multiple phases are consumed and formed at each stage. 

Importantly, to obtain quantitative kinetic parameters from measurements, the full sample system must be measured as function of time. Hence we produced sample pellets to fit entirely within the integrated volume of the expanded 68 keV X-ray beam. We also aligned the lower boundary of the X-ray to the base of the pellets because we observed significant sample shrinkage during experiments. While the quantitative kinetic parameters for these measurements are beyond the scope of this work, our setup lays the experimental foundation for fundamental sample studies in dynamic processing conditions, which can find use in other applications like catalysis.

\subsection{Reversible temperature-induced phase transition of single crystals}
This section presents a sample experiment to demonstrate this chamber's utility in studying a phase transition in single crystal samples under vacuum. \ce{WTe_2} is a quasi-2D material that has a phase transition from the room temperature orthorhombic phase (\ce{T_d}) to the monoclinic (1T') phase, which can be driven by temperature \cite{tao2020t}, laser excitation \cite{sie2019ultrafast}, or high pressure \cite{zhou2016pressure}. The transition has been observed at \textasciitilde565K, though different transition temperatures have been reported in single crystal samples compared to polycrystalline powder samples \cite{tao2020t}. Existing measurements have employed \emph{in-situ} neutron scattering and \emph{in-situ} powder XRD to study this phase transitions. However, to our knowledge, a study utilizing such a broad $q$ range as available in this work has not been previously conducted \cite{tao2020t}.

\begin{figure*}[ht]
\centering
\mbox{\includegraphics[width=\linewidth]{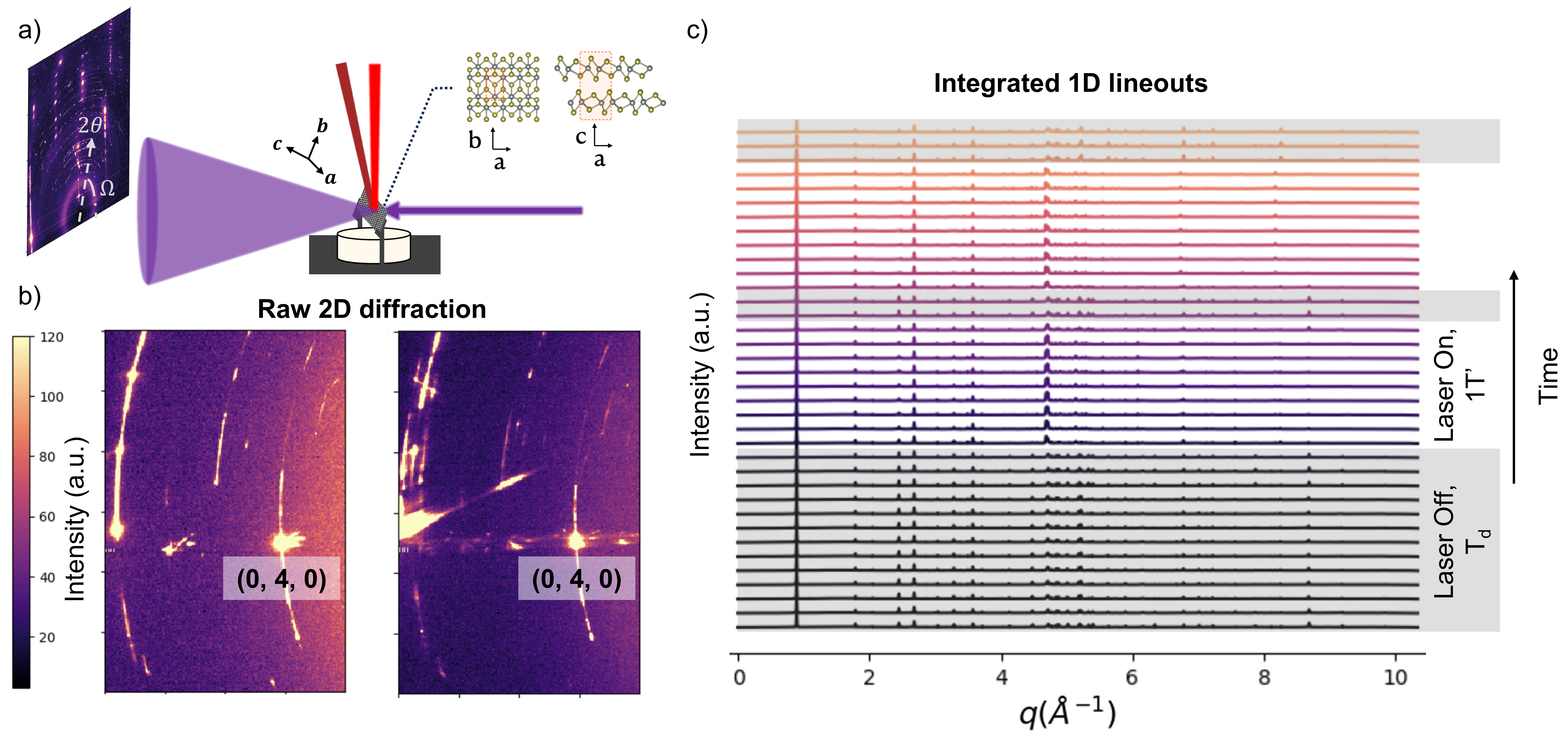}}
\caption{(a) Schematic of the \ce{WTe_2} stacking planes and mounting geometry with respect to the X-ray for a tilted sample. We present the single crystal diffraction data for a sample where $\Vec{b}$ is parallel to the X-ray beam, to clearly showcase the transition during laser on and laser off in: (b) the 2D diffraction and (c) the integrated 1D diffraction data. Here the 1D diffraction data is in a waterfall format to clearly resolve the transition stages.}
\label{fig:combined-wte2}
\end{figure*}

In this work, we used the 976 nm laser to thermally drive the \ce{T_d}-to-1T' transition. Samples were prepared by chemical vapor transport (CVT) with subsequent exfoliation, and were \textasciitilde1 mm thick in the $\Vec{c}$ direction. Enabled by this setup and the 68 keV X-ray energy, we oriented the sample to transmit partially or entirely along $\Vec{b}$, which allowed for the peaks most sensitive to the phase transition to meet the diffraction condition (\ref{fig:combined-wte2}a). XRD measurements along this axis are uncommon due to the material's high X-ray attenuation and challenges with cleaving bulk samples to reduced thicknesses for sufficient signal.  

We observed evidence of a phase transition in the crystal diffraction pattern (Fig. \ref{fig:combined-wte2}b and \ref{fig:combined-wte2}c), made possible by the combination of the heating laser, vacuum environment, high X-ray energy, and broad $q$ range. As this transition has not been driven by laser heating in previous literature, there is uncertainty around the transition temperature. However, this experiment provides the framework for high temperature heating under vacuum for single crystal samples.

\section{Discussion}
The reaction chamber developed in this work enabled 1-s resolved measurements of phase changes induced by reactive gas and laser-heating. By using a laser as the heating source and using remote sensors, we achieved rapid heating rates without compromising measurements due to high temperatures. 1500\degree C surface temperatures were detected with the pyrometer, in various environments: \ce{H_2}, Ar, and vacuum. Future users may change the NIR laser for other wavelengths or powers by adjusting the optics accordingly. The current layout sets the laser and pyrometer to be aligned on a sample surface, regardless of stage motion to change the XRD collection location.

As demonstrated above, the capabilities of this chamber offer two modalities. Quantitative phase analysis was demonstrated with the 1 mm iron oxide pellets that fit entirely within the beam for the duration of the experiment. This was enabled by Rietveld refinement analysis and by the fact that conservation of iron could be assumed. The other modality is to study the sample heterogeneity using a beam smaller than the sample size and that is focused on a selected region with respect to the laser position.

Future upgrades of this setup are ongoing to further enhance the types of materials dynamics that can be studied with this chamber. The most imminent of these are detector upgrades that may enable sub-second measurements. To enable the timing precision for those types of sub-second measurements, we are working to synchronize the laser with the existing X-ray/pyrometer measurements accordingly. Additional efforts are working to incorporate a vacuum-compatible $xyz$-motorized sample stage and partial pressure probes for reactive gasses (e.g. \ce{O_2}, \ce{H_2}, etc.).

\section{Conclusion}
In this work, we have presented a multi-purpose and adaptable reaction chamber for transmission-mode XRD. The chamber capabilities include controlled gas flow up to 5 slm, vacuum pressures down to 0.04 Torr, remote temperature sensing, and up to 10 W of laser power with 1-s XRD acquisition times. To demonstrate the setup, we presented NIR laser-driven heating measurements in two sample types: 1) powder iron oxide reduction with \ce{H_2} and 2) single crystal \ce{WTe_2} under vacuum. Experiments reached at least \ce{1500\degree C} and the associated phase transformation and chemistry as described above. The reaction chamber setup described above expands opportunities to study high-Z and thick samples \emph{operando} and \emph{in situ} at the XPD (NSLS-II 28-ID-2) hutch. 

\section*{Acknowledgments}
The chamber development and primary work for this project (L.M., P.J., S.P., R.G., and L.D.M.) was supported by the U.S. Department of Energy (DOE), Office of Science, Office of Basic Energy Sciences Separation Science, under Award DE-SC0024326. The data collected in Fig. 4 was funded in part by the Advanced Research Projects Agency-Energy (ARPA-E), U.S. DOE, under Award Number DE-AR0001910. The views and opinions of authors expressed herein do not necessarily state or reflect those of the United States Government or any agency thereof. Data associated with Fig. 5 and contributions from S.I. were supported by the DOE, Office of Science, Basic Energy Sciences, Materials Sciences and Engineering Division, under Contract No. DE-AC02-76SF00515. L.Y.M. was supported in part by the National Institute of Standards and Technology (NIST) Graduate Student Measurement Science and Engineering Fellowship. Y.W. was supported by the Stanford Energy Postdoctoral Fellowship. Work by A.A.M. and J.T.M. was supported by the University of Texas at Austin College of Natural Sciences Freshman Research Initiative.

This research used resources and the 28-ID-2 beamline of the National Synchrotron Light Source II, a U.S. DOE Office of Science User Facility operated for the DOE Office of Science by Brookhaven National Laboratory under Contract No. DE-SC0012704.  

We acknowledge Limelight Steel (Andy Zhao, Robert Niederriter, Lwin DuMont) for their helpful discussions and early-stage support and Felipe de Quesada for providing \ce{WTe_2} samples.

\section*{Author contributions}
L.Y.M., L.D.M., and S.G. conceived and designed the research project.
L.Y.M. designed the setup.
L.Y.M., S.P., and P.J., built and tested the prototype.
L.Y.M., S.G., J.T., H.Z., and J.B. implemented the setup at the synchrotron.
A.A.M. and J.T.M. fabricated and provided iron oxide samples.
L.Y.M., P.J., Y.W., S.P., S.I., Z.T., A.A.M., S.G., H.Z., J.T., and J.B. performed experiments at the synchrotron.
L.Y.M., Y.W., R.G., authored the Python scripts to process and analyze the synchrotron results.
L.Y.M. and R.G. analyzed the iron oxide XRD results with Rietveld refinement. 
L.Y.M., S.I., and L.D.M. wrote the paper. All authors discussed and contributed to the paper.

\section*{Data availability}
The raw data presented in this work are available at \url{https://drive.google.com/drive/folders/1UPx2exhB2RwePGqRC3scFO-sBXupoMUF?usp=sharing} and analyzed by code available at \url{https://github.com/leoradm/NSLS-II-Methods-2025} (Dresselhaus-Marais, 2025).

\bibliography{bibliography}

\newpage
\section{Appendix}
\subsection{Viewport windows}
\label{subsection:app1}
Depending on the user's experimental needs, a user might want to replace the window to reduce the XRD background / signal attenuation, find a compatible window when the laser or pyrometer operation wavelength is changed, or reach lower vacuum pressures. The VC22FL and VC23FL viewport mounts allow the windows to be switched out on-site without replacing a full flange with a pre-mounted window. We used 0.5 mm thick $\times$ 1 inch diameter fused quartz for the X-ray entry window and 1 mm thick $\times$ 1.5 inch diameter for the exit window. We calculate 1-mm thick \ce{SiO_2} has 72\% transmittance at 68 keV. One can replace the windows for a different material and thickness as needed, so long as the window will withstand the internal chamber pressure during experiments. One should consult the following equation, which governs the minimum window thickness required to withstand the pressure differential between atmosphere and vacuum:
\begin{align*}
    t = \sqrt{1.1\cdot P\cdot r^2\cdot SF/MR}
\end{align*}
where $P$ is the pressure difference in psi, $r$ is the unsupported radius of the window, $SF$ is the safety factor---typically between 4 and 6---and $MR$ is the modulus of rupture in psi. One should also consider the X-ray attenuation through both X-ray windows, the sample, and the sample holder (if applicable), depending on the sample composition and magnitude of microstructural changes that are to be measured. The maximum window thickness that can be supported by the 1 inch and 1.5 inch viewport mounts are 5 mm and 4 mm, respectively. Since our windows were thinner than the space available in the viewport mounts, we 3D-printed spacer rings to maintain a gas-isolated system inside the chamber.

\end{document}